\documentclass[fleqn,usenatbib]{mnras}

\usepackage{newtxtext,newtxmath}
\usepackage{enumitem}

\usepackage[T1]{fontenc}

\DeclareRobustCommand{\VAN}[3]{#2}
\let\VANthebibliography\thebibliography
\def\thebibliography{\DeclareRobustCommand{\VAN}[3]{##3}\VANthebibliography}
\DeclareMathOperator{\eu}{e}

\usepackage{graphicx}	
\usepackage{amsmath}	
\usepackage{orcidlink}
\usepackage{multirow}

\title[Atmospheric escape under sub-Alfvénic interactions]{Atmospheric escape in hot Jupiters under sub-Alfvénic interactions}

\author[A.~Presa, F.~A.~Driessen and A.~A.~Vidotto]{
Andrés Presa \orcidlink{0009-0006-0590-9346} \thanks{E-mail: presa@strw.leidenuniv.nl },
Florian A. Driessen \orcidlink{0000-0003-3005-7377},
Aline A. Vidotto \orcidlink{0000-0001-5371-2675}
\\
Leiden Observatory, Leiden University, PO Box 9513, 2300 RA, Leiden, The Netherlands\\
}

\date{Accepted XXX. Received YYY; in original form ZZZ}

\pubyear{2015}

\begin{document}
\label{firstpage}
\pagerange{\pageref{firstpage}--\pageref{lastpage}}
\maketitle

\begin{abstract}
Hot Jupiters might reside inside the Alfvén surface of their host star wind, where the stellar wind is dominated by magnetic energy. The implications of such a sub-Alfvénic environment for atmospheric escape are not fully understood.
Here, we employ 3-D radiation-magnetohydrodynamic simulations and Lyman-$\alpha$ transit calculations to investigate atmospheric escape properties of magnetised hot Jupiters.
By varying the planetary magnetic field strength ($B_p$) and obliquity,
we find that the structure of the outflowing atmosphere transitions from a magnetically unconfined regime, where a tail of material streams from the nightside of the planet, to a magnetically confined regime, where material escapes through the polar regions. 
Notably, we find an increase in the planet escape rate with $B_p$ in both regimes, with a local decrease when the planet transitions from the unconfined to the confined regime. 
Contrary to super-Alfvénic interactions, which predicted two polar outflows from the planet, our sub-Alfvénic models show only one significant polar outflow. In the opposing pole, the planetary field lines connect to the star. Finally, our synthetic Ly-$\alpha$ transits show that both the red-wing and blue-wing absorptions increase with $B_p$.
Furthermore, there is a degeneracy between $B_p$ and the stellar wind mass-loss rate when considering absorption of individual Lyman-$\alpha$ wings. This degeneracy can be broken by considering the ratio between the blue-wing and the red-wing absorptions, as stronger stellar winds result in higher blue-to-red absorption ratios. We show that, by using the absorption ratios, Lyman-$\alpha$ transits can probe stellar wind properties and exoplanetary magnetic fields.      
\end{abstract}




\section{Introduction}
Atmospheric escape processes can shape the evolution of a planet, which has major implications for its habitability \citep{Lingam18,owen18,gronoff20}. Moreover, atmospheric escape is also thought to explain certain characteristics in the observed population of short-period exoplanets, such as the scarcity of exoplanets with
radii between 1.5 and 2 Earth radii, known as the ``radius gap'' \citep{Fulton17}, and the lack of planets between 0.03 and 0.3 Jupiter mass, also referred as the ``Neptunian desert'' \citep{Mazeh16}. In the context of atmospheric escape, the presence of a large-scale magnetic field has been traditionally considered to be a protection for planetary atmospheres against atmospheric loss \citep[e.g.][]{lundin07}. However, some recent studies have shown that magnetic fields can enhance certain types of escape. For terrestrial planets, \cite{Gunell18} combined empirical measurements at Venus, Earth, and Mars with semi-analytic models and found that the mass-loss rate could be higher for magnetised planets due to ion outflows through the polar caps and cusps. \cite{Egan19} studied ion escape from a Mars-sized planet using a three-dimensional hybrid simulation, finding that an intrinsic magnetic field does not always cause a decrease in the ion escape rate. Additionally, \cite{Maggiolo22} found that the Earth's magnetic field
enhances the solar energy dissipation in the upper atmosphere, which could increase mass loss. 

There have also been a number of theoretical works investigating how a planetary magnetic field can affect atmospheric escape in close-in exoplanets, particularly hot Jupiters and warm Neptunes. These short-period exoplanets experience substantial irradiation from their host stars, so that photoevaporation is considered to be one of the main drivers of the escape \citep{lammer03,baraffe04,yelle04}. Based on two-dimensional  magnetohydrodynamic simulations, \cite{Khod15} found that the evaporation rate was reduced by
an order of magnitude for a 1\,G field compared to an unmagnetised planet, while \cite{owen14} showed that 
the outflow from the nightside of the planet is greatly suppressed when a magnetic field is included, leading to a net reduction in the escape rate. Recently, some studies have employed detailed 3-D models that account for the planetary magnetic field and the stellar wind as well. Using this kind of simulations, \cite{Arakcheev17} found found a 70 \% reduction in WASP-12’s evaporation rate when the planet’s magnetic field and the stellar wind were included, while \cite{Carolan21} demonstrated that atmospheric escape in magnetised planets takes place through polar outflows, and reported an increase in escape rate as the planetary magnetic field was increased. 

Although for Solar System planets magnetic fields can be determined in situ, for exoplanets, detections are more challenging. 
Some methods used to infer exoplanetary magnetic fields include magnetic star-planet interactions \citep[e.g.][]{Cauley19}, radio emission \citep[e.g.][]{turner21} and spectroscopic transits \citep[e.g.][]{Vidotto10,Kislyakova14,benjaffel22}. 
In relation with the latter method, some earlier studies have analyzed the effects of planetary magnetic fields and stellar winds on the observational signatures of escape, namely the Lyman-$\alpha$ line. Lyman-$\alpha$ transmission spectroscopy provides direct evidence of atmospheric escape, and it has yielded detections in several hot Jupiters \citep{vidalmadjar2003,lecavalierdesetangs2010,lecavalierdesetangs2012,bourrier13} and warm Neptunes \citep{kulow14,Bourrier18}. Other observational signatures, like the  HeI 10830\,\r{A} line, have also been used to capture escaping atmospheres \citep[e.g.][]{allart18,sparke21,zhang2023}. Among the studies that explored how magnetic fields and stellar winds influence these observational signatures, \cite{Trammell11} found increasing transit signal for stronger magnetic fields, while \cite{Villarreal18} showed that the shape of the Ly-$\alpha$ line could be altered by the planetary and stellar magnetic fields, as they determine the size of the magnetosphere and the amount of neutral material retained inside it. Finally, \cite{Carolan21} found increasing line-centre absorption with planetary magnetic field strength, as more low-velocity neutral hydrogen gets trapped in the closed magnetic field lines.

A common aspect of all the works mentioned above is that they consider a super-Alfvénic interaction between the planet and the stellar wind. In this type of interaction, the magnetised wind in which the planet is embedded is super-Alfvénic, i.e., the magnetic energy is smaller than the kinetic energy in the reference frame of the planet. This scenario can be found in all of the magnetised planets of the Solar System, where the interaction between the solar wind and the planetary magnetosphere produces a bow-shock upstream of the planet \citep{Bagenal13}. However, some hot Jupiters could exhibit sub-Alfvénic star-planet interactions due to their close proximity to the host star \citep{Zhilkin19}. In this regime, no bow-shock is formed and the planet interacts with the host star by means of electromagnetic waves. In the Solar System, sub-Alfvénic analogues have been observed in several Jovian satellites \citep{kivelson04} and at Saturn’s satellite Enceladus \citep{dougherty06,pryor11}. Although the theoretical aspects of this kind of star-planet or planet-moon interactions have been extensively studied \citep[e.g.][]{saur13,fischer22}, the impact of magnetic fields on the escaping atmosphere and its spectroscopic signatures within a sub-Alfvénic regime are less explored.

In this work, we employ a three-dimensional, self-consistent radiation-magnetohydrodynamic model to investigate how the planetary magnetic field strength and topology can affect the dynamics of the outflowing atmosphere, the evaporation rates and the resulting Ly-$\alpha$ absorption signals in such a sub-Alfvénic regime. We describe the numerical model in Section \ref{sec:methods}. In Section \ref{section3}, we discuss the effects of the planetary magnetic field on the structure of the escaping atmosphere, while in Section \ref{sec:mass-loss} we analyze the planetary mass loss rate properties. The implications of this results on the Lyman-$\alpha$ transit signatures are presented in Section \ref{sec:spectra}. We discuss our results in Section \ref{sec:discussion}, and present our concluding remarks in Section \ref{sec:conclusions}.

\section{3-D RADIATION-MAGNETOHYDRODYNAMIC MODEL OF ATMOSPHERIC ESCAPE}\label{sec:methods}
To model the escaping atmosphere of a magnetised planet embedded in a magnetised stellar wind, we employ the self-consistent, radiation-magnetohydrodynamic code presented in \cite{Carolan21}. This model assumes that the
planetary atmosphere is fully composed of ionized and neutral hydrogen and solves the following ideal magnetohydrodynamic (MHD) equations:

\begin{equation}
\frac{\partial{\rho}}{\partial{t}} +\nabla \cdot \rho {\mathbf{ u}} = 0,
\end{equation}
\begin{align}
\label{eqn:momentum}
\frac{\partial(\rho\mathbf{u})}{\partial t} + \nabla \cdot \Bigg[\rho \mathbf{u} \mathbf{u} + \left(P + \frac{B^2}{8\pi}\right)I - \frac{\mathbf{B}\mathbf{B}}{4\pi}\Bigg] = & \nonumber \\
\rho \bigg( \mathbf{g} - \frac{GM_{*}}{|\mathbf{R}|^2} \hat{R} 
- \mathbf{\Omega} \times (\mathbf{\Omega} \times \mathbf{R}) & - 2(\mathbf{\Omega} \times \mathbf{u}) \bigg),
\end{align}

%
\begin{align}
\label{eqn:energy}
\frac{\partial e}{\partial t} + \nabla \cdot \Bigg[\mathbf{u} \Bigg(e + P + \frac{B^2}{8\pi}\Bigg) - \frac{(\mathbf{u} \cdot \mathbf{B}) \mathbf{B}}{4\pi}\Bigg] = & \nonumber \\
\rho \bigg( \mathbf{g} - \frac{GM_{*}}{|\mathbf{R}|^2} \hat{R} - \mathbf{\Omega} \times (\mathbf{\Omega} \times \mathbf{R}) \bigg) \cdot \mathbf{u} & + {\cal H} - {\cal C},
\end{align}
%
\begin{equation}
    \frac{\partial\mathbf{B}}{\partial t} + \nabla \cdot (\mathbf{u}\mathbf{B} - \mathbf{B}\mathbf{u}) = 0,
\end{equation}
where $\rho$, $\mathbf{u}$, $P$, $\mathbf{B}$ and $I$ are the mass density, velocity, thermal pressure, magnetic field and identity matrix, respectively. Furthermore, $\mathbf{r}$ is the position vector relative to the centre of the planet, $\mathbf{a}$ is the orbital distance, $\mathbf{\Omega}$ is the orbital rotation rate and  $\mathbf{R} = \mathbf{r} + \mathbf{a}$ is the position vector relative to the star. The total energy density $e$ is given by 
\begin{equation}
e = \frac{\rho u^2}{2} + \frac{P}{\gamma -1} + \frac{B^2}{8\pi},    
\end{equation}
where $\gamma=5/3$ is the adiabatic index. The centrifugal force
$\mathbf{\Omega} \times (\mathbf{\Omega} \times \mathbf{R})$ and the Coriolis force $2(\mathbf{\Omega} \times \mathbf{u})$ are included in the momentum equation (\ref{eqn:momentum}), in addition to the planetary and stellar gravities. 

In equation \eqref{eqn:energy}, the change
of internal energy of the atmosphere is balanced by
the heating and cooling terms $\mathcal{H}$ and $\mathcal{C}$. The volumetric heating rate due to  stellar radiation is
\begin{equation}
   {\cal H} = \eta \sigma n_n F_{\rm xuv}\eu^{-\tau},  
\end{equation}
where $\eta$ is the excess energy released after a hydrogen atom is ionised, $\sigma$ is the ionization cross-section of the hydrogen atom, $n_n$ is the number density of neutrals, $F_{\rm xuv}$ is the incident XUV flux and $\tau$ is the optical depth for neutral hydrogen. Following \cite{Carolan21,Hazra22}, we assume that the incident stellar radiation is plane parallel, and that the entire XUV spectrum is
concentrated at 20\,eV. This corresponds to $\sigma$ = $1.89\times10^{-18}$ cm$^{-2}$  and $\eta$ = 0.32. The incident XUV flux is injected into the grid from the negative $x$ direction, so that the optical depth can be calculated at any distance $x$ from the left (i.e., negative-$x$) edge of the grid $x_{\rm left}$ as
\begin{equation}
    \tau(x) = \int_{x_{\rm left}}^{x}n_n \sigma \text{d}x.
\end{equation}
The energy conservation equation \eqref{eqn:energy} also contains a volumetric cooling rate ${\cal C}$, which is the sum of cooling due to emission of Ly-$\alpha$ radiation \citep{osterbrock89}
\begin{equation}
{\cal C}_{{\rm Ly}\alpha}= 7.5 \times 10^{-19} n_p n_n \eu^{(-1.183 \times 10^5/T )},
\end{equation} 
and the cooling due to collisions with free electrons \citep{Voronov97}:
\begin{equation}\label{eqn:cooling}
 {\cal C}_{\rm col} = 2.91 \times 10^{-8} n_e n_n \frac{U^{0.39}}{0.232+U}\eu^{-U} \chi_H, \; U=\frac{1.58\times 10^5}{T}. 
\end{equation}
Here, $T$ is the temperature, $\chi_H = 2.18\times 10^{-11} $erg is the ionisation potential of hydrogen, and $n_p$ and $n_e$ are the number density of  protons and electrons. All the above equations use cgs units. 

\begin{table*}
    \centering
    \caption{The stellar and planetary parameters used in our models. $M_{p}$ and $R_{p}$ are the mass and radius of the planet; $B_{p}$ is the polar dipole field strengths and obliquity describes the magnetic field obliquities; $a$ is the orbital distance from the star, and $F_{\text{xuv}}$ is the X-ray flux received by the planet. The stellar mass and radius are denoted by $M_*$ and $R_*$, respectively. $\dot{M}_*$ and $T_*$ describe the stellar wind mass-loss rate and temperature, and $B_{*}$ is the stellar surface field strength (radial).
}
    \begin{tabular}{cccccccccccc}
        \hline
        $M_{p}$ & $R_{p}$ & $B_{p}$& Obliquity & $a$ & $F_{\text{xuv}}$ & $M_*$ & $R_*$ & $\dot{M}_*$ & $T_*$ & $B_{*}$\\
        ($M_J$) & ($R_J$) & (G)& ($^\circ$) &(au) & (erg/cm$^2$/s) & ($M_\odot$) & ($R_\odot$) & ($M_\odot/\text{yr}$) & ($10^6$\,K) & (G)\\
        \hline
        0.7 & 1.4 & 0.25$-$25 & 0, 45, 90, 135, 180 & 0.05 & 500 & 1.148 & 1.19  & 0, $2\times 10^{-13}$  & 1 & 2\\
        \hline
    \end{tabular}
    \label{tab:model_parameters}
\end{table*}

Our model also solves two additional equations to track the density of neutrals and ions:
\begin{equation}
   \frac{\partial n_n}{\partial t} + \nabla \cdot n_n\mathbf{u} = \mathscr{R} - \mathscr{I},
\end{equation}
\begin{equation}
   \frac{\partial n_p}{\partial t} + \nabla \cdot n_p\mathbf{u} = \mathscr{I} - \mathscr{R},
\end{equation}
where $\mathscr{I}$ denotes the ionisation rate due to photoionisation \citep[e.g.,][]{MurrayClay09} plus electron-impact ionisation \citep{Voronov97}
\begin{equation}\label{eqn:ionization}
 \mathscr{I} =\frac{ \sigma n_n F_{\rm xuv} \eu^{-\tau}}{{h\nu}} + 2.91 \times 10^{-8} n_n n_e \frac{U^{0.39}}{0.232+U}\eu^{-U}.
\end{equation}
Here, $U$ has the same form as in equation \eqref{eqn:cooling}. Last, the recombination rate is \citep{Storey95}
\begin{equation}\label{eqn:recombination}
 \mathscr{R} = 2.7 \times 10^{-13} (10^4/T)^{0.9} n_e n_p.
\end{equation}
Both $\mathscr{R}$ and $\mathscr{I}$ are given in cgs units  (cm$^{-3}$ s$^{-1}$). Finally, we assume that all particles are in thermal equilibrium with each other and have a common plasma temperature. 

To perform our atmospheric escape simulations we solve the equations described previously using the \textsc{bats-r-us} code \citep{toth12,gombosi21}. Specifically, the ideal MHD equations are solved using a second-order numerical scheme with Linde flux limiter and minmod slope limiter to calculate the Riemann Problem fluxes at cell interfaces. The time step is advanced using an explicit, second-order temporal scheme and is further constrained by setting the Courant-Friedrichs-Lewy (CFL) number to 0.8. 

We perform our simulations in a 3-D Cartesian cube with $x,y,z \in[-40,40]R_{p}$, where the orbital plane is in the $xy$-plane and the planet is centered in the origin of the coordinate system. We use a minimum cell size of 1/16\,$R_{p}$ within a radius of 5\,$R_{p}$, which gradually increases to a maximum cell size of 2\,$R_{p}$ at the edges of the grid. With these static grid refinements, this results in 5.4 million cells in total. 
For simplicity, we assume that the planet is tidally locked to its host star, which is located outside the numerical domain in the $-x$ direction.

We impose an inner domain at the surface of the planet (1\,$R_{p}$), where we fix the base temperature to 1000\,K and the base density to $2.4\times10^{11}$ cm$^{-3}$ based on 1-D steady-state models \citep{Allan19}. Regarding the velocity, the material starts at nearly zero speed at 1\,$R_{p}$. In order to remove any unwanted inflow near the outer boundaries of the grid, we use inflow limiting boundary conditions in the outer boundaries, with the exception of the $-x$ boundary of the stellar wind, which has an inflow boundary condition \citep{mccann19,carolan21sw,Carolan21}.

The simulation is initialized with a spherically symmetric velocity profile of the escaping atmosphere, which is a fit to a 1-D model from \citet{Allan19}. In this work, the profile adopts the form $u_r = u_\infty (1-R_p/r)^\beta$, where $u_\infty=43$ km/s is the terminal velocity of the outflow and $\beta = 3.66$ was found to be the best fit to the 1-D model. Our model was chosen to represent a hot Jupiter similar to HD\,209458 b.

We treat the magnetic field of the planet as a dipole, whose desired polar strength is obtained at 1\,$R_{ p}$. At the planetary surface, a floating boundary condition is applied, such that the field lines can adjust to changes in the outflow properties. This is included in the  \textsc{bats-r-us}’ Global Magnetosphere module default boundary.

The stellar wind is radially injected at the negative $x$-boundary. Here, we use the same boundary conditions as in \cite{Carolan21}, and provide values of the stellar wind
velocity, temperature, density and magnetic field taken from an external 1-D isothermal stellar wind model \citep{Johnstone15}, whose parameters are given in Table \ref{tab:model_parameters}. The stellar magnetic field is radial with positive polarity, i.e., the magnetic field lines are directed outwards from the star. 
Additionally, we fix the
stellar wind temperature and mass loss rate $\dot{M}_*$ at the boundary of the computational domain. Together with the velocity solution of the 1-D model $\mathbf{u}_{\mathrm{sw}}=(u_r,0,0)$, the stellar mass-loss rate sets the density at the boundary of the simulations: $\rho_{\mathrm{sw}} = \dot{M}_*/(4\pi R^2 u_r)$. Once the stellar wind
solution is found at each point of the $-x$ boundary, we keep these values fixed throughout the computation to simulate a steady inflow. 
In this work, the stellar wind is sub-Alfvénic at the planet's orbit, in contrast with the super-Alfvénic simulations performed in \cite{Carolan21}. This allows us to investigate another possible regime of the stellar wind and its implications for the planetary outflow.   

\section{Structure of the planetary outflow}\label{section3}
\begin{figure*}
    \centering
    \includegraphics[width=0.33\textwidth]{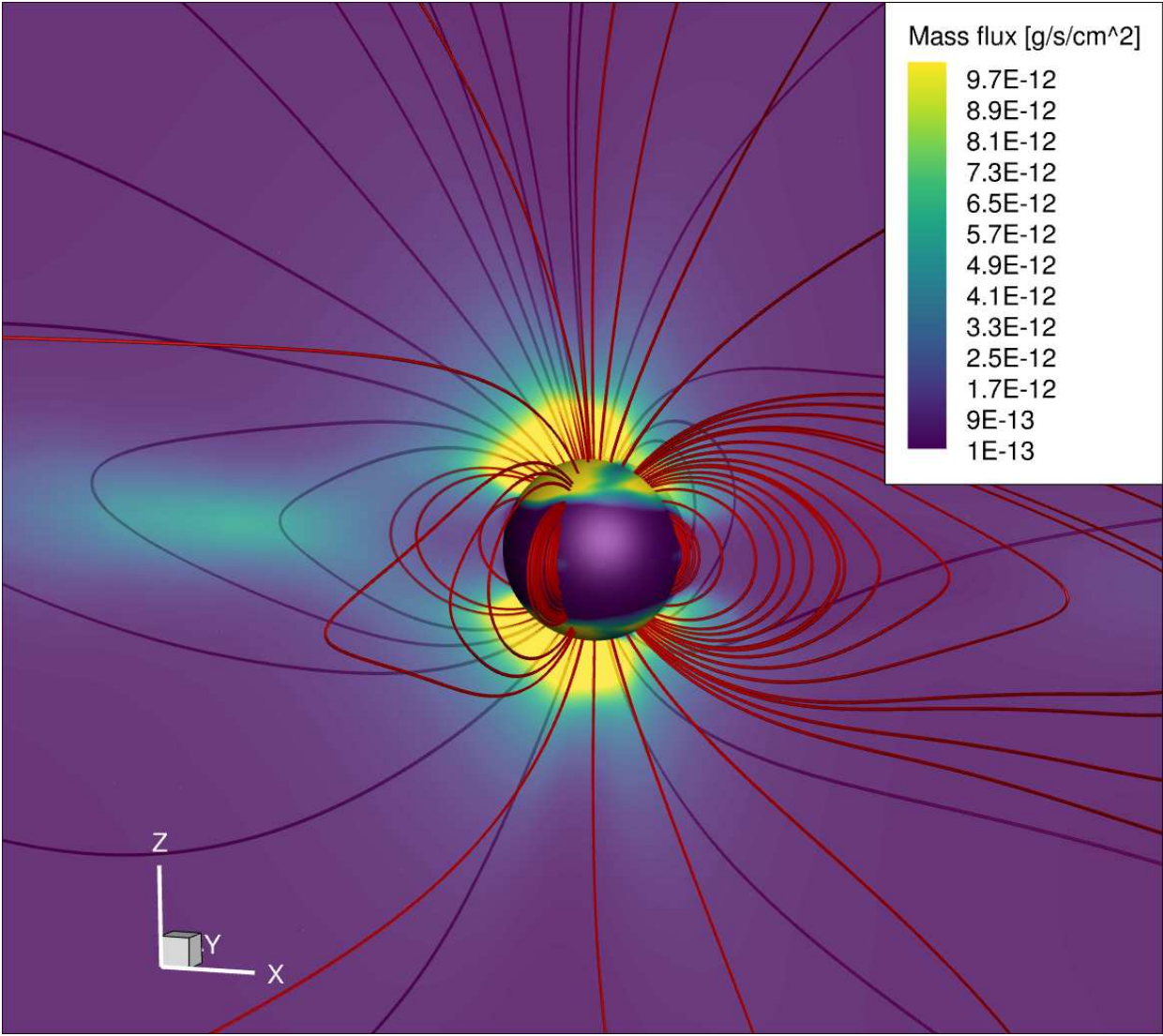}
    \includegraphics[width=0.33\textwidth]{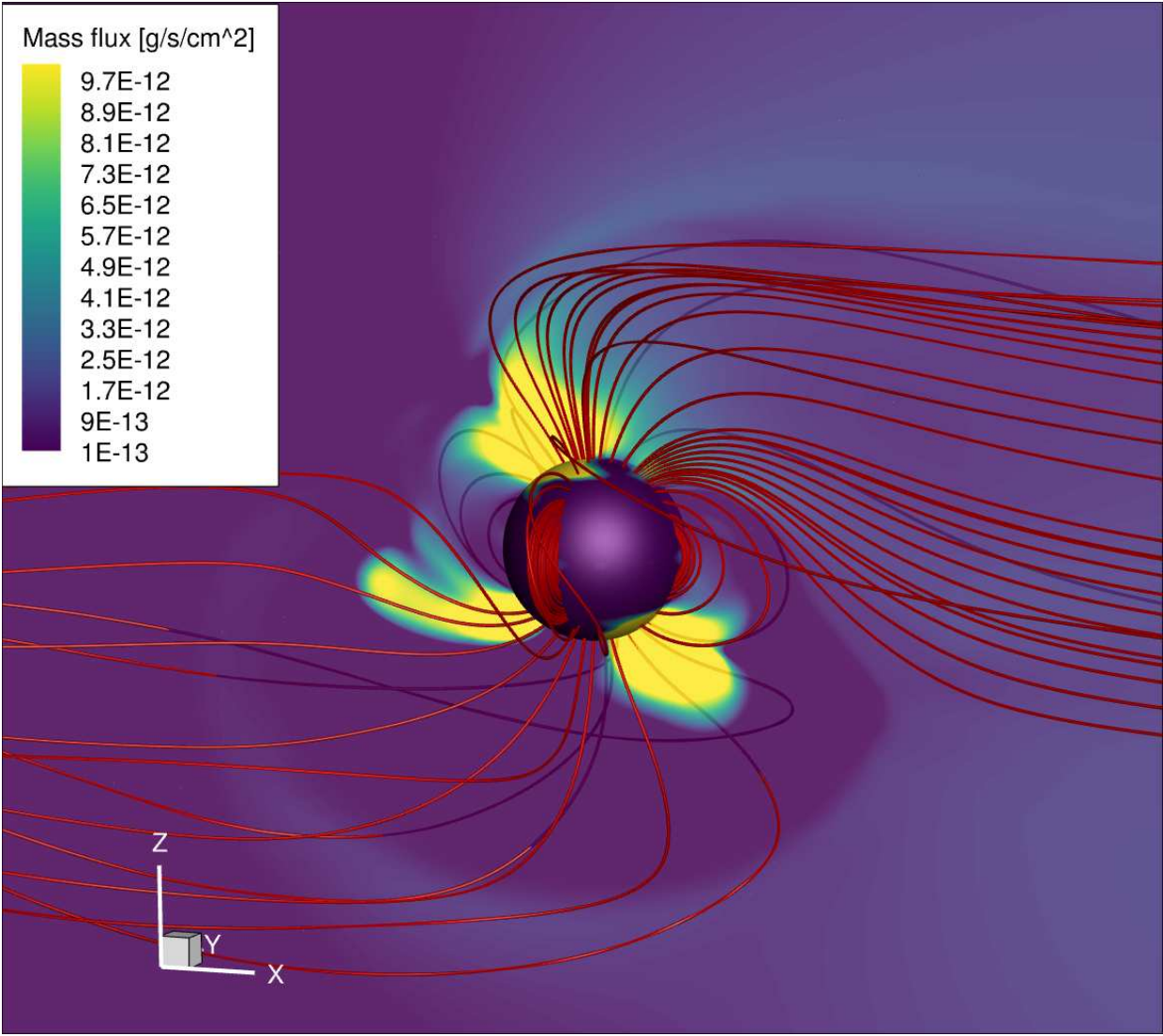}
    \includegraphics[width=0.33\textwidth]{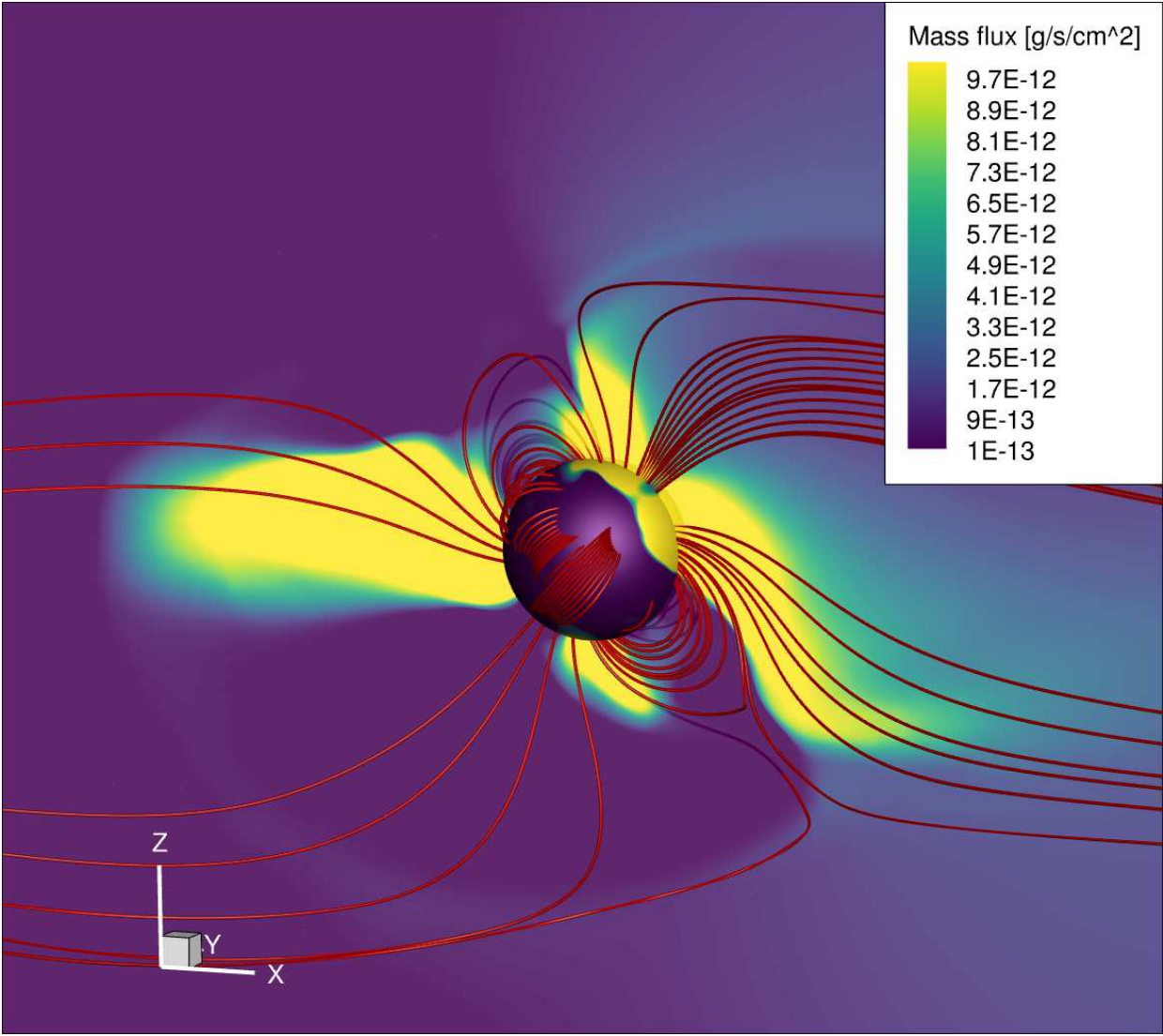}
    \caption{Polar-plane cut of the mass flux of the evaporating planetary atmosphere for models with $B_p=25$\,G. The red lines are the magnetic field lines of the planet (central sphere). Left: model with no stellar wind, 0$^\circ$ obliquity. Middle: model with stellar wind, 0$^\circ$ obliquity. Right: model with stellar wind, 45$^\circ$ obliquity. }
    \label{fig:3DFigures}
\end{figure*}

We construct a grid of models using the parameters reported in Table \ref{tab:model_parameters}. In particular, we vary the dipolar field strength of the planet, ranging from 0.1\,G to 25\,G at the poles., and the magnetic obliquity, varying from 0$^\circ$ to 180$^\circ$ in  45$^\circ$ increments with respect to the rotational axis of the planet (additional obliquities are presented for the 5-G model in Section \ref{sec:discussion_tilt}). This results in 40 models, each examined under two scenarios, leading to a total of 80 models: (i) a case with no stellar wind, 
and (ii) a case with a stellar wind that is ten times stronger than the solar value of $2\times10^{-14}$\,$M_\odot/{\rm yr}$. Finally, we evolve each of these models until a steady-state solution is found, where there are no large-scale variations with additional computational time and no large-scale variations occur anymore in the model.
Figure \ref{fig:3DFigures} shows the mass flux of the planetary escaping atmosphere resulting from three of such simulations.

The magnetic field properties of hot Jupiters are still poorly understood. By modelling observational features, some studies suggested that the magnetic fields of close-in exoplanets range from 10 to 120~G \citep[e.g.,][]{Vidotto10, Cauley19, benjaffel22}, with other studies suggesting sub-G field strengths \citep{Kislyakova14}. 
Our grid of models incorporate fields within this observed range. Planetary dynamo theories take into account 
the complex interplay between internal composition, heat distribution, geological activity, rotation, and core dynamics. Some early theoretical studies suggested that the surface magnetic field of close-in planets would have fields that are weaker than Jupiter's due to their slower rotation caused by tidal locking \citep[e.g.][]{griessmeier04}. However, more recent works suggest that the magnetic dynamo is less affected by rotation, but instead is driven by the heat flux in gas giants, thus predicting surface magnetic fields of $\sim10 - 100$\,G \citep{christensen09,yadav17}. These results are in line with recent observational estimations \citep{Cauley19}.

Planetary magnetic fields can also exhibit different topologies. 
In the solar system, Earth, Jupiter, Uranus and Neptune all exhibit dipolar tilts \citep{Bagenal13}, resulting in non-axisymmetric magnetic fields. However, advanced dynamo simulations are also capable of producing nearly axisymmetric magnetic fields, as observed in Saturn \citep{yadav22}. In the case of close-in exoplanets, which are likely in tidally-locked orbits,  it is unclear whether they would exhibit significant magnetic obliquities. Although stellar magnetic dynamo processes are different from those of planets, it is worth noting that some synchronized close binary stars exhibit substantial dipolar inclinations \citep[e.g.][]{
kochukhov19, tsvetkova24, hahlin24}. 
Due to these uncertainties in the geometry of planetary magnetic fields, we consider a sufficiently broad parameter space in order to explore a variety of potential dipolar tilts. While in our study we only consider dipolar fields, theoretical studies suggest that some planets could also exhibit multipolar fields \citep{gastine2012}.

\subsection{Atmospheric escape models with no stellar wind}

\begin{figure*}
    \centering
    \includegraphics[width=0.98 \textwidth]{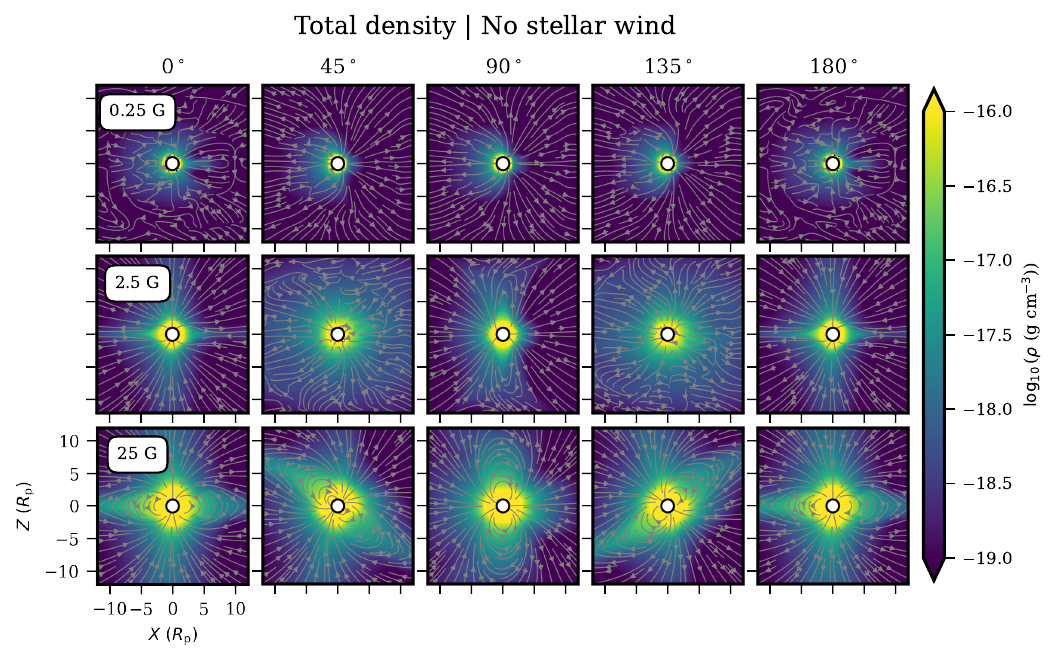}
    \includegraphics[width=0.98\textwidth]{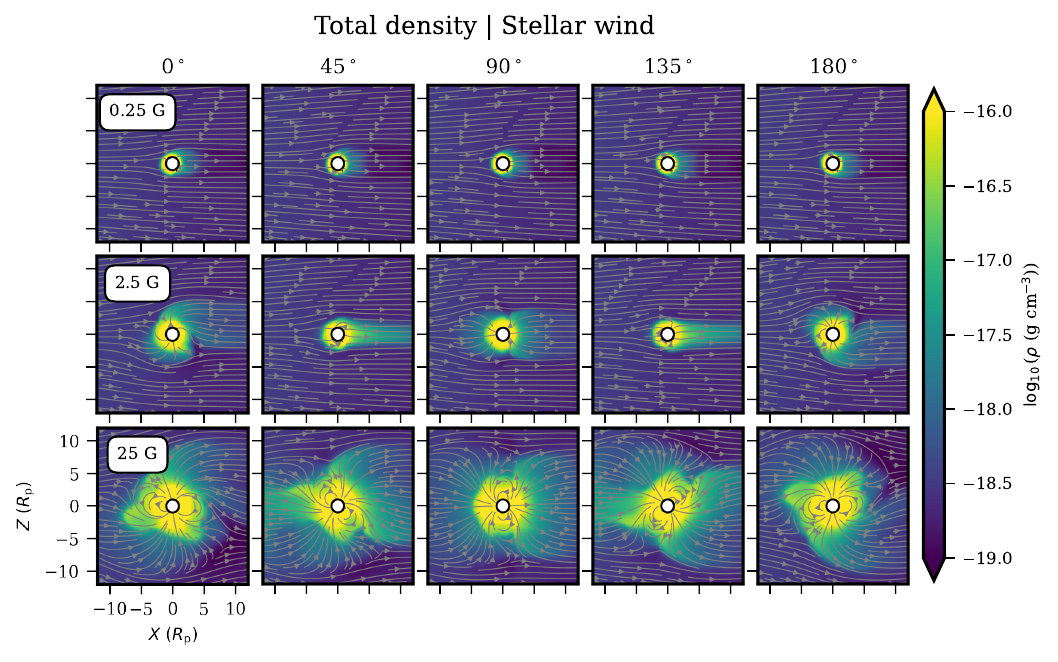}
    \caption{Slice along the polar plane for selected models. Each row corresponds to a different magnetic field strength of the planet, while each column represents a different magnetic obliquity. The colors indicate the density structure around the planet, and the grey streamlines trace the magnetic field lines. (Top) models with no stellar wind. (Bottom) models with a stellar wind mass-loss rate of $2\times10^{-13}$\,$M_\odot/{\rm yr}$.}
    \label{fig:density}
\end{figure*}

\begin{figure*}
    \centering
    \includegraphics{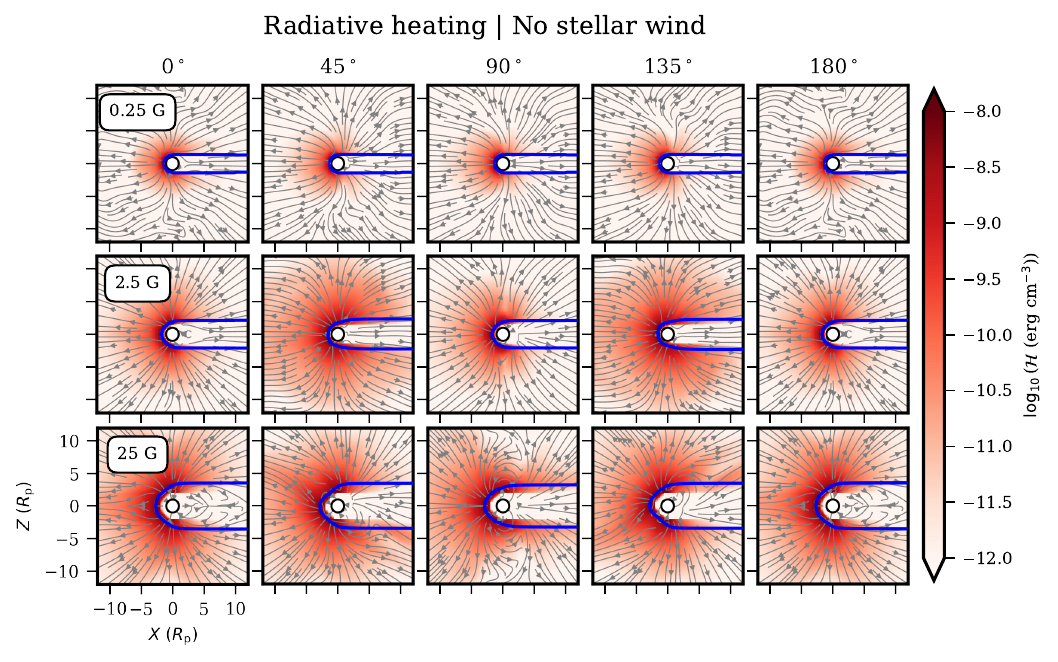}
    \includegraphics{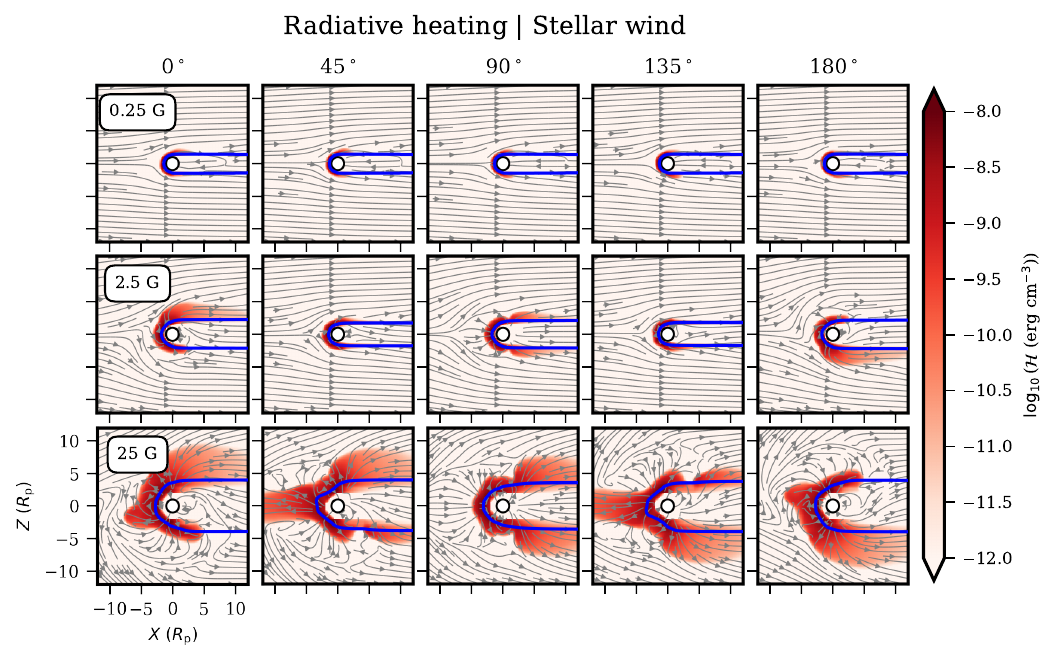}
    \caption{Similar to Figure \ref{fig:density}. Here, the color represents the radiative volumetric heating rate around the planet, while the grey streamlines trace the velocity flow in each model. The blue lines mark the points where the optical depth $\tau$ for the Lyman-$\alpha$ photons reaches unity. }
    \label{fig:heating}
\end{figure*}

Figures~\ref{fig:density} and \ref{fig:heating} show the total density and the volumetric heating rate along the polar plane for a selection of models, respectively.
The top set of panels illustrates the models with no stellar wind, while the bottom set of panels displays those with stellar wind. Within each panel set, the rows represent different strengths of the planetary magnetic field, and the columns correspond to different magnetic obliquities.
In the absence of a stellar wind, the magnetic field lines are open when the planetary magnetic field is weak, as the pressure of the planetary flow dominates over the magnetic pressure. Similar to models of unmagnetised planets, tidal effects from the star divert the planetary outflow from being radial, with tidal forces contributing to accelerate material away from the planet on the dayside \citep[e.g.][]{Carroll17}. 

As the planetary magnetic field strength increases, a well defined magnetic structure starts to develop. Once a closed magnetosphere is formed, atmospheric escape occurs via two polar outflows above and below the orbital plane, where the field lines are predominantly open. Within the closed-field regions, a deadzone of high density, low velocity material forms. This can be seen in the left panel of Figure \ref{fig:3DFigures} and in the 25\,G panels of Figure \ref{fig:density} (denoted by ``No stellar wind''). This behaviour has been found in previous works \citep[e.g.][]{Khod15,Arakcheev17,Carolan21}. We also observe an increase in the heating rate around the planet as the planetary dipolar field $B_{p}$ becomes stronger. This is because a larger magnetosphere traps more material inside it, which results in more photoionization and a more extended heating profile. Consequently, the $\tau=1$ surface (blue lines in Figure \ref{fig:heating}), where a significant amount of photons are deposited, shifts away from lower to higher altitudes as well. We note that the heating rate remains essentially zero in the wake of the planet, since the nightside is not exposed to the stellar radiation. 

Furthermore, the shift from a magnetically unconfined to a magnetically confined regime is not solely controlled by the strength of the planetary magnetic field; the magnetic obliquity also has a significant impact on the transition between these types of structures. This can be observed in Figure \ref{fig:density} for a 2.5\,G magnetic field strength: with obliquities of 45$^\circ$ and 135$^\circ$, the flow blows open most field lines, while other obliquities already show closed loops. As we will discuss in Section \ref{sec:mass-loss}, this obliquity effect has implications for the total mass-loss rate of the planet. 

\subsection{Atmospheric escape models including stellar wind}
The cases involving the stellar wind are depicted in the lower sets of panels of Figures \ref{fig:density} and \ref{fig:heating}. Stellar winds can exert pressure confinement around the expanding atmosphere of the planet \citep{vidotto2020}, preventing part of the planetary atmosphere from escaping. This phenomenon is indeed seen in our models with weak planetary magnetic fields. Similar to the work by \cite{carolan21sw}, we observe a suppression of the dayside outflow, with material redirected towards a tail in the wake of the planet. Similar to the models with no stellar wind, a magnetosphere develops and  grows in size as $B_{p}$ increases (see Table \ref{tab:magnetosphere}), and the $\tau=1$ surface moves outward from the planet. The magnetospheric size $R_{m}$ is given in Table \ref{tab:magnetosphere} for the $0^\circ$ obliquity model. In super-Alfvénic conditions, $R_m$ is estimated by balancing the ram pressure of the stellar wind with the magnetic pressure of the planet. Because our models are in the sub-Alfvénic regime and the stellar wind kinetic energy is low, its ram pressure is several orders of magnitude smaller than its thermal pressure. Therefore, we estimate this distance as the point in the magnetic equator where the magnetic pressure of the planet matches the thermal pressure of the stellar plasma. As expected, $R_{m}$ increases with $B_p$. 

\begin{table}
    \centering
    \caption{The size of the planet's magnetosphere $R_{m}$ along the magnetic equator for the models with stellar wind, chosen as the point where the magnetic and thermal pressures are balanced. This is illustrated here for a magnetic obliquity of 0$^\circ$, with very similar values obtained for the other magnetic obliquities.}
    \begin{tabular}{|c|c|}
        \hline
        $B_p$ (G) &$R_{m}$ ($R_{p}$) \\
        
        \hline
        2.5 & 2.9 \\
        
        5 & 3.8 \\
        
        10 & 4.8 \\
        
        25 & 6.3 \\
        \hline
    \end{tabular}
    \label{tab:magnetosphere}
\end{table}

An important difference between our magnetically confined models and previous MHD models found in the literature \citep{Khod15,carolan21sw,Carolan21}
is the absence of a bow shock around the planet. This is because the stellar wind in our models is sub-Alfvénic at the orbital location of the planet. The implications of a sub-Alfvénic regime of the stellar wind are more pronounced for stronger planetary magnetic fields, where the magnetic pressure dominates over the ram and thermal pressures further away from the planet. In such cases (see the 2.5\,G and 25\,G models in the lower panels of Figures \ref{fig:density} and \ref{fig:heating}), we observe a tail of material steaming from the north magnetic pole of the planet, where the magnetic field lines are open. Conversely, at the south magnetic pole, the stellar and planetary magnetic field lines connect, and no such tail originates from the south pole. Instead, some planetary material flows back to the star from the vicinity of the Lagrange L1 point, along the field lines of the stellar wind, closely resembling the \textit{sub-Alfvénic quasi-closed envelope} scenario reported by \cite{zhilkin20}. We discuss in more detail the differences between sub-Alfvénic and super-Alfvénic models and their implications in Section \ref{sec:discussion}.

The magnetic obliquity of the planet determines the position of its magnetic poles, influencing the connection between the planetary and stellar wind field lines and, consequently, the strength of the dayside outflow. In our models, we observe the most significant stream of material for a 45$^\circ$ and 135$^\circ$ obliquity, as can be seen in Figures \ref{fig:3DFigures} (middle and left panels) and \ref{fig:density}. For other obliquities, the dayside outflow is confined by the stellar wind much closer to the planet. Similar to the no-stellar-wind scenario, the planetary magnetic obliquity also influences the tail structure. For $B_\text{p}=2.5$\,G, the outflow is in the magnetically confined regime for obliquities of 0$^\circ$, 90$^\circ$, and 180$^\circ$, while it is in the unconfined regime for obliquities of 45$^\circ$ and 135$^\circ$.

\section{Atmospheric mass-loss rate properties}\label{sec:mass-loss}
In the previous section, we discussed the planetary atmosphere morphology arising from different stellar wind and planetary magnetic field conditions. To investigate the differences in escape rate among these models, we compute the planet's total evaporation rate by integrating the mass flux ($\rho \mathbf{u}$) through a closed surface area $A$ around the planet
\begin{equation}\label{eq:mass_loss}
    \dot{M}=\oint_A \rho \mathbf{u} \cdot {d}\mathbf{A}.
\end{equation}
The mass flux of the planet on the polar plane can be seen in Figure \ref{fig:3DFigures} for three of our models. We choose to integrate the mass flux through a cube around the planet, where each side of the cube is 30\,$R_{p}$. We experimented with cubes of different dimensions, yielding very similar mass-loss rates in all cases. 

\subsection{Escape rate dependence on magnetic field strength}
\begin{figure*}
    \centering
    \includegraphics[width=1\textwidth]{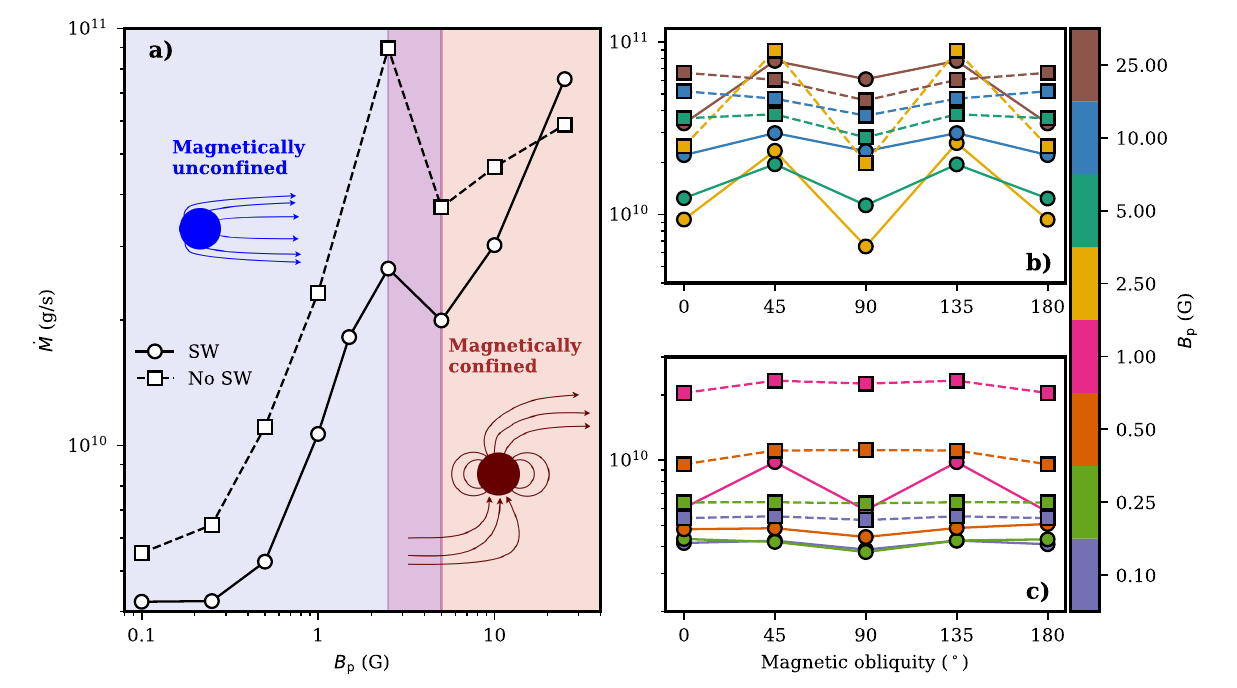}
    \caption{\textbf{a)} Atmospheric mass-loss rate as a function of the planetary magnetic field for a magnetic obliquity of 45$^\circ$. The circles and solid lines represent the models with stellar wind, and the squares and dashed lines represent the models with no stellar wind. \textbf{b), c)} Atmospheric escape rate as a function of the planetary magnetic obliquity. Here, the models are split according to their magnetically confined (b) and unconfined (c) regimes.
    The points are color-coded according to $B_p$. Circles and squares have the same meaning as in panel a).}
    \label{fig:mdot}
\end{figure*}

Figure \ref{fig:mdot}a shows the total atmospheric escape of the planet for different magnetic field strengths and a fixed magnetic obliquity of 45$^\circ$. Here, we choose to show this case as it represents an intermediate configuration between a dipole aligned with the polar plane (0$^\circ$) and a dipole aligned with the orbital plane (90$^\circ$). Nevertheless, we find qualitatively equivalent results for the other obliquities (see Appendix \ref{appendix:mdot}). Our simulations show a non-monotonic increase in mass-loss rate with magnetic field strength. The evaporation rate increases in both the magnetically unconfined and confined regimes, but decreases when the planet transitions between these regimes. This phenomenon can be understood as follows: 
\begin{itemize}[leftmargin=*]
    \item[-] In the magnetically unconfined regime ($B_{p}\leq2.5$\,G) the thermal and ram pressures of the escaping material dominates over the magnetic pressure around the planet, and the field lines are predominantly open. In this case, the magnetic field is not able to trap almost any material. On the contrary, ionized particles experience a $\frac{1}{c}(\mathbf{j}\times\mathbf{B})$ Lorentz force directed away from the planet ($\mathbf{j}$ is the current density and $c$ the speed of light). Since the magnitude of this force increases with $B_p$, this leads to more escape as the dipolar field strengthens.
    \item[-] As the planetary magnetic field strength increases, the planet transitions from a magnetically unconfined regime, described above, to a magnetically confined one. In the latter scenario, the planet exhibits an extended magnetosphere, with closed loops around the equator and open field lines at the poles. In the equator, the magnetic forces trap particles inside the close-field lines, and the material is constrained to escape through the poles. Consequently, we observe a mass-loss rate reduction in the transition region between both regimes ($2.5$\,G\,$\leq B_{p}\leq$ 5\,G).
    \item[-] Once the planet is in the magnetically confined regime, the magnetosphere expands as the dipolar field increases (see Table \ref{tab:magnetosphere}). This expansion results in a growing high-density region of both neutral and ionized material surrounding the planet, as seen in Figure \ref{fig:density}. As a result, the stellar XUV photons encounter more XUV-absorbing material  as they head towards the planet. This leads to increased radiative heating around the planet (see Figure \ref{fig:heating}), and a higher rate of atmospheric escape. Here, it is important  to note also that when energy and momentum deposition takes place in the sub-critical part of the outflow, as is the case in our models, there is an increase in mass-loss rate. This is similar to the theory of stellar winds \citep[e.g.,][]{lammer03}. 
\end{itemize}

Overall, we find the total mass-loss rate can vary by more than an order of magnitude between 0.1\,G and 25\,G in the two stellar wind conditions studied in this work. However, we note that our simulations show a reduction in planetary escape rates in the stellar wind scenario for most models compared to the no-wind scenario (see Figure \ref{fig:mdot}a). This reduction is more evident in cases with weak and moderate planetary magnetic fields, where the stellar wind inhibits the dayside outflow. This is a similar behavior as that studied in non-magnetised models \citep{vidotto2020, 2020MNRAS.498L..53C}.
Conversely, in the magnetically confined regime, the dayside arm extracts a substantial amount of material due to the connection between stellar and planetary magnetic field lines. 
As a result, when the magnetic structure is closed, the increase in mass-loss rate with $B_{p}$ is more pronounced in the stellar wind scenario. For the highest magnetic field strength considered in this study, 25\,G, the mass-loss rate is already higher in the stellar wind simulations than in the no stellar wind ones for obliquities of 45$^\circ$ and 90$^\circ$.

\subsection{Escape rate dependence on magnetic obliquity}
The atmospheric escape rate trends with varying magnetic field strengths are similar across all magnetic obliquities considered in this study. However, the local decrease in mass loss rate occurs at different magnetic field strengths depending on the magnetic obliquity of the planet. For a magnetic obliquity of 45$^\circ$ (Figure \ref{fig:mdot}a), this decrease is observed between 2.5\,G and 5\,G. In contrast, for obliquities of 0$^\circ$ and 90$^\circ$, the decrease occurs between 1.5\,G and 2.5\,G (see Appendix \ref{appendix:mdot}). This variation is due to the dependence of the transition point between unconfined and confined magnetic field structures on the planet's magnetic obliquity (c.q.~Section \ref{section3}).

The dependence of atmospheric escape on magnetic obliquity is illustrated in Figures \ref{fig:mdot}b and \ref{fig:mdot}c for various magnetic field strengths. For $B_{p}=2.5$ G, the 45$^\circ$ model shows up to $\sim$3 times more escape than the other obliquities. As mentioned previously, the 2.5\,G model is in the open regime for a 45$^\circ$ obliquity, while it has already transitioned to the closed regime for the other obliquities.

In general, the impact of magnetic obliquity is more significant for strongly magnetised planets exposed to a substantial stellar wind. In these conditions, we observe a peak escape rate at obliquities of 45$^\circ$ and 135$^\circ$, with a minor increase of up to a factor of two compared to the aligned models. The 45$^\circ$ and 135$^\circ$ tilt models have a geometry that favors the connection of planetary field lines with the stellar field lines, increasing the dayside outflow towards the star (see right panel of Figure \ref{fig:3DFigures}). In comparison, the dayside outflow stagnates closer to the planet for the other models.

\section{Absorption profiles}\label{sec:spectra}

To assess the observational implications of our results, we simulate the spectroscopic transit using the Ly-$\alpha$ line profile of each model at mid transit. We do so by using a ray-tracing model originally developed by \citet{Vidotto18} and adapted by \citet{carolan21sw} to perform these calculations on our 3-D grids. Firstly, the ray-tracing model calculates the line-of-sight (LOS) velocity as $u_{\mathrm{los}}=-u_x+y\Omega$, where $\Omega$ is the orbital rotation rate. Given the density of neutrals from our simulations, the frequency-dependent optical depth is given by:
\begin{equation}\label{eq:tau}
    \tau_\nu=\int n_n \sigma \phi_\nu \text{dx}, 
\end{equation}
where $\sigma=\pi e^2 f/ (m_e c)$ is the frequency-dependent absorption cross section at line centre, $f=0.416410$ is the quantum mechanical oscillator strength for the Ly-$\alpha$ transition, $m_e$ the electron mass, $e$ the electron charge, and $c$ the speed of light. We take the line profile to be a Voigt profile, which accounts for both Doppler and natural line broadening. 

Once the optical depth is found, the fraction of transmitted intensity can be found as
\begin{equation}\label{intensity}
    \frac{I_\nu}{I_*}=\eu^{-\tau_\nu},
\end{equation}
where $I_*$ is the emitted specific intensity from the star and $I_\nu$ is the specific intensity attenuated by the absorption from the planet and its atmosphere. We assume a uniform specific intensity over the stellar disc at a given frequency (i.e., no limb darkening).  The fraction of specific intensity absorbed by the planet is therefore $1-I_\nu/I_*$. The transit depth can then be calculated as:
\begin{equation}\label{eq:transit_depth}
    \Delta F_\nu = \frac{ \int \int (1-\eu^{-\tau_\nu})\text{d}y\text{d}z}{\pi R^2_*},
\end{equation}
where d$y$ and d$z$ define an element of area associated to each ray in our simulation. We consider a uniform grid with 201 cells in each coordinate direction of the plane of the sky, and calculate the transit depth in 51 velocity bins between $-$500 and 500 km/s. The frequency-dependent rays are aligned plane-parallel to the star-planet line (x-axis), and the line profiles were calculated at mid-transit, for an impact parameter $b=0$.

\begin{figure}
    \centering
    \includegraphics[width=0.5\textwidth]{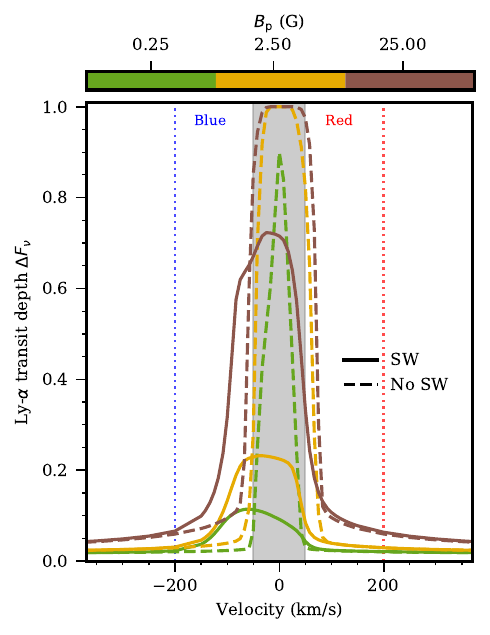}
    \caption{Transit depth of the Ly-$\alpha$ line at mid-transit for models with three magnetic field strengths with stellar wind (solid lines) and no stellar wind (dashed lines), assuming 0$^\circ$ magnetic obliquity. The blue and red vertical lines denote the outer bounds for the integrals of the blue and red wing absorptions considered in Figure \ref{fig:percentage_absorption}, while the gray shaded area represents the line centre.  }
    \label{fig:transit_depth}
\end{figure}

The transit depths of several selected models are shown in Figure \ref{fig:transit_depth}. Similar to \citet{Carolan21}, we find more absorption in the line centre as the magnetic field of the planet increases. This is caused by the low velocity neutral material trapped in the closed field lines around the planet,  as seen in Figure \ref{fig:density}. As the size of the deadzone grows with the planetary magnetic field strength, so does the absorption in the line centre. We also observe increasing blue-wing absorption with $B_{p}$, since the outflow is more directed along the line-of-sight towards the observer for stronger magnetic fields. These trends are present in the models irrespective of the strength of the stellar wind.

\begin{figure*}
    \centering
    \includegraphics{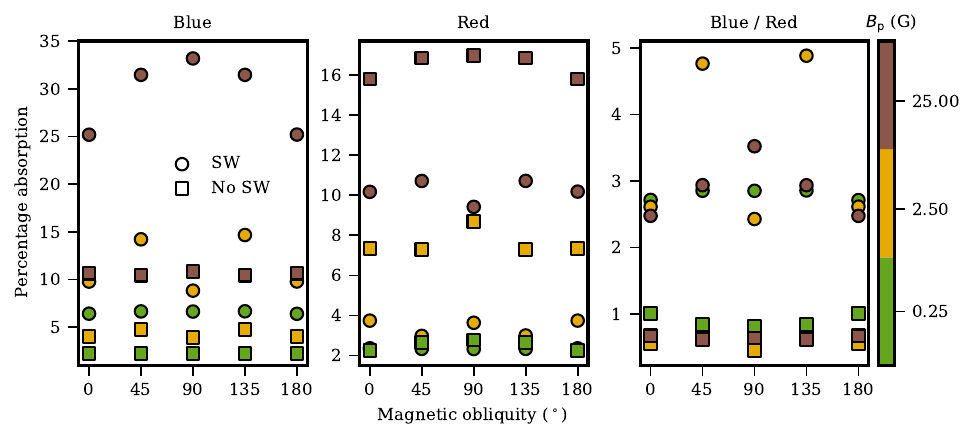}
    \caption{Integrated Ly-$\alpha$ absorption in the blue  ($-$200 km/s to $-$50 km/s, left) and red  (50 km/s to 200 km/s, middle) wings, and their blue-to-red ratios (right) Circles depict models with stellar wind and squares with no stellar wind. Notice that different models that show similar absorption in one of the wings can be distinguished from each other by looking at their respective blue-to-red ratios.}
    \label{fig:percentage_absorption}
\end{figure*}

Fixing the planetary magnetic field strength, we observe a more asymmetric transit in the stellar wind scenario. The stellar wind pushes the neutral material in the direction of the observer, resulting in excess absorption in the blue wing and less absorption in the centre of the line, compared to the no stellar wind case. These features can be examined quantitatively in Figure \ref{fig:percentage_absorption}, where we show the percentage absorption in the blue ($-$200 km/s to $-$50 km/s) and red (50 km/s to 200 km/s) wings for a variety of magnetic field strengths and obliquities in each stellar wind scenario. As mentioned before, we observe more blue-wing absorption as the magnetic field of the planet increases, due to more material being accelerated through the polar outflows and towards the observer. The red-wing absorption also grows with the dipolar field strength, since some of this material travels back in the direction of the star along magnetic field lines away from the observer. For weak and moderate planetary fields, the absorption signatures are very similar for every magnetic obliquity. For a strong dipolar field and a typical stellar wind, we see that the blue-wing absorption percentage increases nearly 10\,\% from 0$^\circ$ to 90$^\circ$ obliquities. This is because the magnetic poles move closer to the orbital plane as the dipole is tilted, and the polar outflows are more funnelled towards the LOS. These features can also be observed in the line profiles, where we appreciate a stronger signal in the blue wing and the line centre for the strongest $B_p$ as the magnetic obliquity increases from $0^\circ$ to $90^\circ$  (see Appendix \ref{appendix:signatures}).

Finally, we show the ratio between blue-wing and red-wing absorption in the right panel of Figure \ref{fig:percentage_absorption}. Under the no-stellar wind condition (squares), this ratio remains around $0.5-1$ for all the planetary magnetic field strengths considered. For some models, there is slightly more absorption in the red wing due to the effect of the orbital and tidal forces. Conversely, the models including stellar wind show several times more absorption in the blue wing than in the red wing. As previously mentioned, this blueshift results from the planetary material being pushed to the tail by the stellar wind. As we will discuss in Section \ref{sec:discussion}, these ratios can be used to constrain stellar wind properties and planetary magnetic fields.   

\section{Discussion}\label{sec:discussion}

\subsection{Different types of outflow dynamics }
Based on our results, and on previous studies that investigated a super-Alfvénic type of interaction, we can now identify three main outflow structures in hot Jupiter planets depending on the characteristics of their magnetic field and their interaction with the stellar wind: 
\begin{itemize}[leftmargin=*]
    \item[-] Unmagnetised planets exhibit a comet-like tail of material centered at the orbital plane, trailing the planet as it orbits the star. If the stellar wind is very weak, similar to our no stellar wind simulations, material also escapes through the dayside due to the tidal action of the star. For stronger stellar winds, the dayside material is redirected towards the planetary tail, and the dayside outflow is reduced or even suppressed \citep{carolan21sw}. 
    This situation is also seen in weakly magnetised planets, where the pressure associated with the planetary magnetic field is not strong enough to control the outflow dynamics.
    \item[-] Magnetised planets in the super-Alfvénic zone of the stellar wind display a double-tail structure, where two polar ouflows produce two tails of escaping material \citep{Carolan21}. These tails exhibit some degree of asymmetry due to the wind-planet magnetic interaction. In addition, a deadzone appears around the planet, consisting of material trapped within the closed magnetic field lines \citep{owen14,Khod15,Carolan21}.
    \item[-] With radial stellar wind topology, magnetised planets in the sub-Alfvénic regime of the stellar wind present a polar outflow at one of the poles of the planet. At the other magnetic pole, planetary field lines connect to the star and no such tail is found, albeit in some instances, material could move towards the star. In this regime, the planet also features a deadzone of trapped material surrounding the equator.  
\end{itemize}

\begin{figure*}
    \centering
    \includegraphics[width=0.99\textwidth]{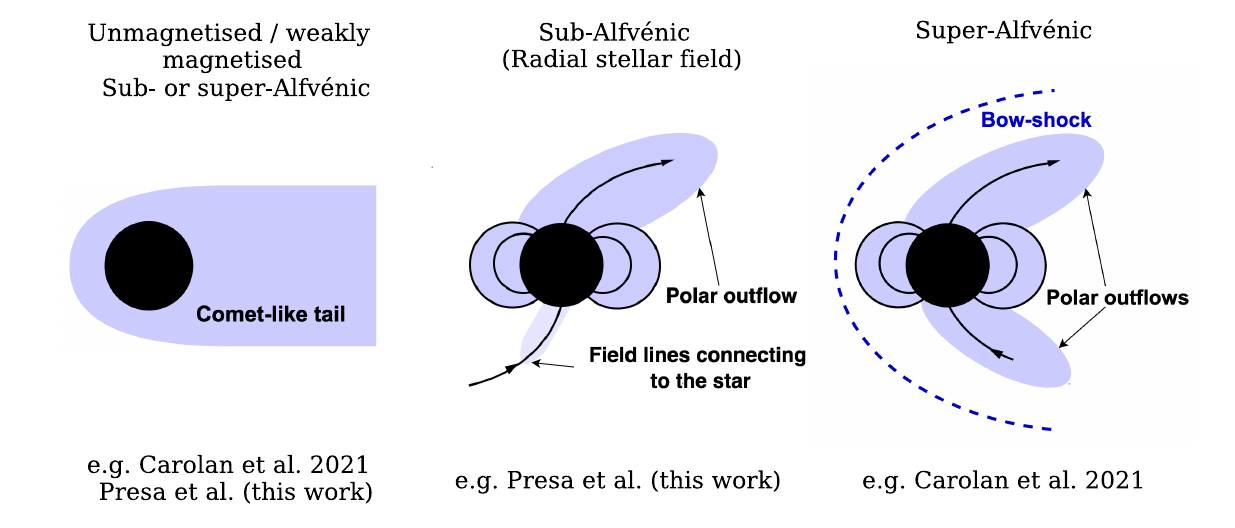}
    \caption{Schematic of the three main atmospheric escape dynamic structures. (Left) Unmagnetised or weakly magnetised planet. (Middle) Magnetised planet under a sub-Alfvénic interaction with the stellar wind. (Right) Magnetised planet under a super-Alfvénic interaction with the stellar wind.}
    \label{fig:dynamic_cases}
\end{figure*}

A cartoon of the three types of dynamic structures is given in Figure \ref{fig:dynamic_cases}. This description complements the classification of the possible types of magnetospheres of hot Jupiters described in \cite{Zhilkin19}. Namely, the subtypes \textit{B1} (\textit{shock-less intrinsic magnetosphere}) and \textit{A1} (\textit{intrinsic magnetosphere with bow shock}) reported in that work correspond to the sub-Alfvénic and super-Alfvénic regimes mentioned here.

\subsection{Influence of the planetary magnetic field on the mass-loss rate}

In this work, we find that the total mass-loss rate can increase by more than one order of magnitude with the increase of $B_p$ from 0.1\,G to 25\,G. Our models show that the evaporation rate increases with the planet's magnetic field strength even in the magnetically confined regime, where the planet has a well formed, extended magnetosphere. As explained in Section \ref{sec:mass-loss}, a larger magnetosphere retains more atmospheric material and, thus, XUV-absorbing atoms, leading to more heating of the upper atmosphere of the planet. In turn, this increases atmospheric escape rates, since the energy budget available to accelerate atmospheric material increases and is deposited in the sub-critical region of the flow. This paradigm, that higher $B_p$ could lead to stronger escape, originally proposed based on  theoretical and empirical considerations for terrestrial planets \citep{blackman18,Maggiolo22}, is observed in our simulations of hot Jupiters, although the physical reasons differ. In the context of hot Jupiters, previous works found that a planetary magnetic field reduced the total amount of escape \citep{owen14,Arakcheev17}, although these studies did not include radiative heating self-consistently. 

Compared to the magnitude of the magnetic field, its orientation has a minor effect on the atmospheric escape rate. The most notable differences are found for a field of 2.5\,G, where the $45^\circ$ model shows three to four times more escape than the other obliquities. As previously mentioned, this depends on whether the planet is within the magnetically unconfined regime or in the magnetically confined regime. In comparison to other works, the calculations in \cite{Zhilkin20_obliquity} show an even weaker dependence on magnetic obliquity, although their study is limited to a single $B_p$ and the interaction is super-Alfvénic. Likewise, \cite{dong19} found that the planetary magnetic obliquity is unlikely to change the long-term atmospheric evolution of an Earth-like planet, as the escape rates varied by about a factor of two at most.   

\subsection{Outflow properties under different magnetic obliquities}\label{sec:discussion_tilt}
Our simulations show that the structure of the escaping atmosphere can share a high degree of similarity for certain pairs of magnetic obliquities. In particular, the models with a magnetic obliquity of 180$^\circ$ display the same structure as the models with a magnetic obliquity of 0$^\circ$ flipped over the orbital plane. This leads to very similar evaporation rates and Ly-$\alpha$ transit signatures at mid disk. However, the observational signatures would be inverted for positive and negative impact parameters (i.e., transits above or below the line with $b=0$), since the polar outflow originates below the orbital plane for a 180$^\circ$ obliquity. An analogous situation occurs between the 45$^\circ$ and 135$^\circ$ models.

\begin{figure*} 
    \centering
    \includegraphics{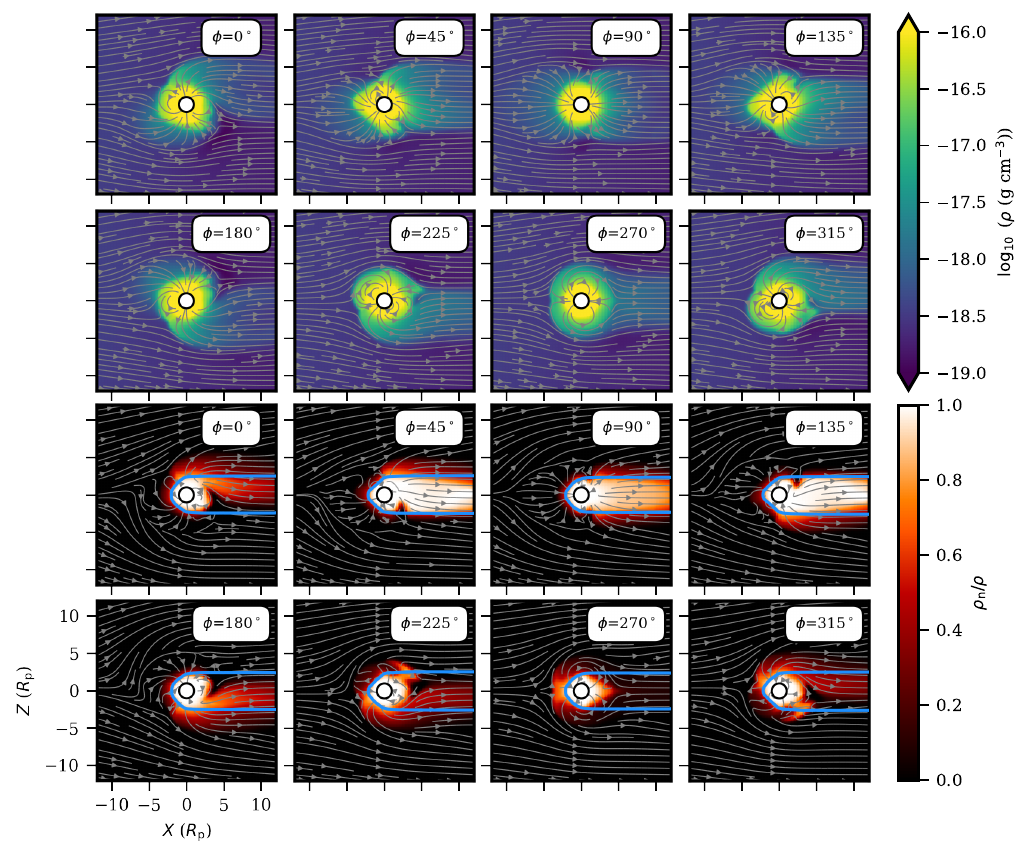}
    \caption{Polar plane cuts for different models with $B_p=5$\,G and different magnetic obliquities ranging from 0$^\circ$ degrees to 315$^\circ$. The first two rows show the total density distribution, and the streamlines trace the magnetic field lines of each model. The last two rows show the density of neutrals normalized to the total density. In this case, the grey streamlines represent the velocity flow around the planet. The blue lines indicate the $\tau=1$ surface.} 
    \label{fig:density_extended}
\end{figure*}

To investigate the complete parameter space of magnetic obliquity, we computed an additional set of models with obliquities of 225$^\circ$, 270$^\circ$ and 315$^\circ$. This was done for a planet with $B_{p}=5$\,G subject to the adopted stellar wind. The total density along the polar plane is shown in the first two rows of Figure \ref{fig:density_extended} for magnetic obliquities ranging from 0$^\circ$ to 315$^\circ$ in increments of 45$^\circ$. We note that the 45$^\circ$ model corresponds to the 225$^\circ$ model with opposite polarity. Likewise, the 90$^\circ$ model corresponds to the 270$^\circ$ model, and the 135$^\circ$ model to the 315$^\circ$, each with an opposite polarity, respectively. Due to the reversed polarity of the planetary field, there is little to no connection between the planetary and stellar field lines on the dayside in the 225$^\circ$, 270$^\circ$ and 315$^\circ$ obliquity models, as opposed to the counterpart models with opposite polarities (45$^\circ$, 90$^\circ$, and 135$^\circ$, respectively). This effect is particularly prominent in the 270$^\circ$ case, where the open planetary field lines are completely diverted towards the wake of the planet by the stellar wind, resulting in a single comet-like tail streaming from the nightside of the planet. This is the only simulation with a closed magnetosphere, in contrast with the quasi-open configurations found in the other models.

\begin{figure}
    \centering\includegraphics{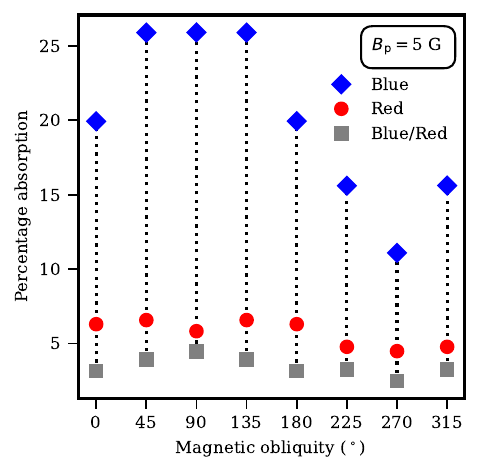}
    \caption{Ly-$\alpha$ absorption in the blue (blue diamonds) and red (red circles) wings, along with their blue-to-red ratios (grey squares) for $B_p=5$\,G and magnetic obliquity ranging from 0$^\circ$ to 315$^\circ$. The integration bounds for the blue and red wing absorptions are the same as those used in Figure \ref{fig:percentage_absorption}.} 
    \label{fig:absorption_extended}
\end{figure}

To assess the impact on the line-profile absorption for each obliquity model, Figure \ref{fig:absorption_extended} shows the computed percentage absorption at mid-transit in both the red and blue wings of the Ly-$\alpha$ line, and their corresponding ratio. The maximum absorption in the blue wing is observed at magnetic obliquities of 45$^\circ$, 90$^\circ$ and 135$^\circ$, while the 225$^\circ$, 270$^\circ$ and 315$^\circ$ models exhibit more than 10 percentage points lower blue-wing absorption. The origin of such differences in absorption lies in the amount of absorbing (neutral) material present in the planetary tail. The last two rows of Figure \ref{fig:density_extended} display the density of neutral hydrogen in the polar plane normalised by the total density. As shown in the figure, in the 45$^\circ$, 90$^\circ$ and 135$^\circ$ models, the tail is primarily made up of neutral material. This is because the bulk of the planetary outflow originates from regions shielded from stellar radiation, as indicated by the $\tau=1$ surface (blue lines in Figure \ref{fig:density_extended}).
In contrast, in the 225$^\circ$, 270$^\circ$ and 315$^\circ$ models the main outflow starts more oriented towards the dayside, where there is substantial deposition of stellar radiation. This outflow is subsequently deflected towards the nightside by the stellar wind.
By the time this material reaches the planetary tail, most of the hydrogen has been ionized. 
As a result, there is less absorbing material funneled towards the line of sight, leading to reduced blue-wing absorption.

In contrast to the blue-wing absorption, Figure \ref{fig:absorption_extended} shows that the red-wing absorption remains relatively stable across all magnetic obliquities at around $5-6\%$. This suggests that the red-wing signal could be better suited for constraining the planetary magnetic field strength, while the blue wing is more sensitive to changes in magnetic obliquity. Furthermore, the blue-to-red ratios are all greater than one, indicating the presence of a substantial stellar wind, as discussed in Section \ref{sec:spectra}. 

Although the above discussion applies only to the 5\,G model, we expect similar trends for the rest of our magnetically confined models, that is, decreasing relative blue-wing absorption and quasi constant red-wing absorption. 

\subsection{Implications for observations: degeneracy between stellar wind properties and planetary field strength}
The results presented in the previous sections show that the strength and geometry of the stellar and the planetary magnetic fields determine to a great extent the dynamics of the planetary outflow and the type of interaction with the star. As shown in Sections \ref{sec:mass-loss} and \ref{sec:spectra}, this can affect planetary mass-loss rates and the observational signatures of atmospheric escape. It has been found that the line centre shows the most univocal signature of a significant planetary magnetic field for both sub-Alfvénic (this work) and super-Alfvénic conditions \citep{Carolan21}, regardless of the strength of the stellar wind. According to MHD models, the line centre shows more absorption than in the pure hydrodynamical case due to the growing deadzone of high-density, low-velocity material trapped inside the closed field loops.   

One problem with Ly-$\alpha$ transits, however, is that the centre of the line is contaminated by the interstellar medium absorption and geo-coronal airglow emission. As a consequence, the most significant absorption signal is typically detected in the Ly-$\alpha$ blue wing \citep{vidalmadjar2003,Ehrenreich15,Lavie17,dosSantos20}. Our results show that characterizing the magnetic field of the planet from blue-shifted signatures alone is challenging, as there is a degeneracy between the planetary magnetic field strength and the stellar wind strength. For example, Figure \ref{fig:percentage_absorption} shows that a 10\% absorption in the blue wing could be explained by either a 25\,G planet under a weak stellar wind or a 2.5\,G planet under stronger stellar wind conditions. 

One way to disentangle this degeneracy is to consider the ratio between the blueshifted and redshifted absorptions. As argued in Section \ref{sec:spectra}, a line profile which shows several times more absorption in the blue wing than in the red wing is indicative of an atmosphere being blown away by a substantial stellar wind. 
Conversely, if this ratio remains close to one, it suggests a much weaker stellar wind mass-loss rate. Other mechanisms that could contribute to the blue-shifted absorption are radiation pressure and charge exchange \citep{Shaik16}, which were neglected in this work.

Our simulations also indicate that both the blue-wing and red-wing absorption increase with the magnetic field strength of the planet. Therefore, the presence of a magnetic field could also explain some unusually high-redshifted signatures, such as the one detected in the warm Neptune GJ\,3470b by \cite{Bourrier18}. 

\section{Conclusions}\label{sec:conclusions}
In this work we investigated how a planetary magnetic field affects the atmospheric escape properties of a hot Jupiter (prototypical of HD\,209458 b) located in the sub-Alfvénic regime of the stellar wind, and compared the results with a no-wind scenario. To do so, we use a similar 3-D self-consistent radiation-MHD model following \cite{Carolan21}. For each stellar wind condition, we vary the magnetic field strength of the planet and its obliquity to examine how these changes affect the dynamic structure of the outflow, the total mass-loss rate of the planet and the observational signatures of the escape. 

We find that the structure of the outflowing planetary material is altered with increasing $B_p$, when the planet shifts from a magnetically unconfined regime to a confined regime. In the former case, the escaping outflow is able to blow open most field lines, whereas in the latter scenario the planet develops a magnetosphere, with closed field lines around the equator and open field lines at the magnetic poles.
In the magnetically confined regime, we show that in sub-Alfvénic conditions only one polar outflow forms, and some field lines connect back to the star in the other pole of the planet. This outcome is contrary to the super-Alfvénic interactions mainly studied in the literature, where two polar outflow form (see Fig. \ref{fig:dynamic_cases} for a sketch of the different atmospheric escape structures). We also found an increase in the total evaporation rate in both the magnetically unconfined and confined regimes. 
In the latter regime, other works found a reduction in the evaporation rates \citep{owen14,Khod15,Arakcheev17}.
We attribute the $\dot{M}$ increase due to an increase in the heating rate in the sub-critical part of the flow  with increasing $B_p$ (i.e., more XUV-absorbing material is trapped in the closed field lines thus leading to higher heating). This result places importance on introducing self-consistent heating and cooling mechanisms in atmospheric escape simulations. In the magnetically unconfined regime, charged particles experience an outward $\frac{1}{c}(\mathbf{j}\times\mathbf{B})$ force that accelerates them away from the planet. This results in increasing escape rates up until the point where the planet transitions to the magnetically confined regime.

Additionally, we investigated how changing the planetary magnetic field strength and orientation affected the synthetic Lyman-$\alpha$ signal for each stellar wind condition. Similar to \cite{Carolan21}, we found an increase in line-centre absorption with increasing $B_p$ due to the growing deadzones around the planet. The blue-wing absorption was found to increase with magnetic field strength, as the amount of material moving towards the observer also increases with $B_p$. Likewise, the red-wing absorption displays a similar trend with magnetic field strength, as more material moves back to the planet along the closed field lines. While the blue-wing absorption can show some variability with magnetic obliquity for a given magnetic field strength, the red-wing absorption remains more stable across different magnetic obliquities. Potentially,  the red-wing signal could be used for constraining the planetary magnetic field strength, while the blue wing could be used to constrain the magnetic obliquity.

Finally, we find that different planetary magnetic fields and stellar wind conditions can yield very similar Lyman-$\alpha$ absorption depths in one of the wings. 
 In this regard, we show that the blue-wing to red-wing absorption ratio can be used to break the degeneracy present in individual Lyman-$\alpha$ wings. This is because the ratio between blue-wing and red-wing absorption increases with the stellar wind mass-loss rate, as more material is funnelled towards the observer by the stellar wind. 
For example, a line profile which shows several times more absorption in the blue wing than in the red wing is indicative of an atmosphere being blown away by a substantial stellar wind. 
Conversely, if this ratio remains close to one, it suggests a much weaker stellar wind mass-loss rate. 
Therefore, Lyman-$\alpha$ transits can be used to constrain stellar wind properties and planetary magnetic fields.
Our work is relevant for the interpretation of atmospheric escape observations, as some poorly understood red-wing signals could be attributed to a planetary magnetic field. Moreover, our work describes the atmospheric escape properties of a planet under a sub-Alfvénic star-planet interaction, a scenario that could be common for close-in exoplanets and resulting star-planet signatures.

\section*{Acknowledgements}

We thank the reviewer for their valuable comments and suggestions, which helped improve the quality of our paper. 
The research performed in this work received funding from the European Research
Council (ERC) under the European Union’s Horizon 2020 research and innovation programme (grant agreement no.~817540, ASTROFLOW). This work used the Dutch national e-infrastructure with the support of the SURF Cooperative using grants no.~EINF-7347 and EINF-5173. This work used the \textsc{bats-r-us} tools developed at the University of Michigan Center for Space Environment Modeling and made available through the NASA Community Coordinated Modeling Center.

\section*{Data Availability}

The data underlying this article will be shared on reasonable request to the corresponding author.



\bibliographystyle{mnras}
\bibliography{bibliography.bib}



\appendix
\section{Mass-loss rates for several planetary magnetic obliquities}\label{appendix:mdot}
\begin{figure*}
    \centering
\includegraphics{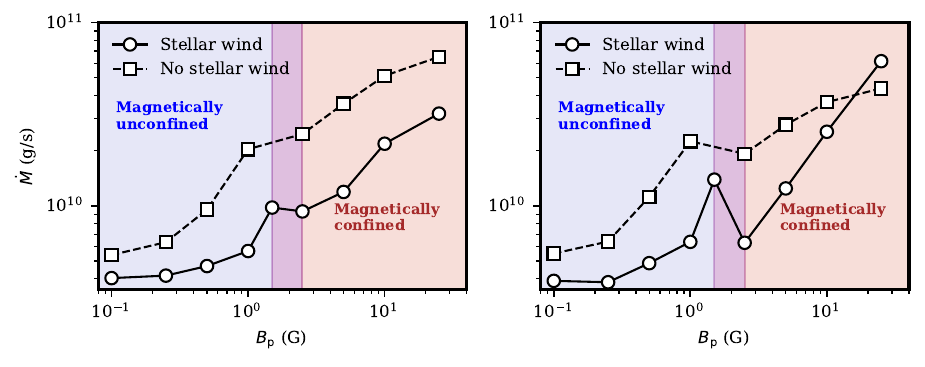}
    \caption{Similar to Figure \ref{fig:mdot}a, for magnetic obliquities of 0$^\circ$ (left) and 90$^\circ$ (right).}
    \label{fig:mdot_appendix}
\end{figure*}
Here, we present the mass-loss rates as a function of $B_p$ for the $0^\circ$ and $90^\circ$ obliquity cases. The corresponding plots are shown in Figure \ref{appendix:mdot}. Similar to Figure \ref{fig:mdot}a, we indicate the magnetically unconfined and magnetically confined regimes according to the dynamic structure of the outflow (see Figure \ref{fig:density}). In both cases, we observe a local decrease in the evaporation rate as the planet transitions from the unconfined regime to the confined regime (1.5\,G $\leq$ $B_p$ $\leq$ 2.5\,G). We point out that this differs from the 45$^\circ$ obliquity models shown in Figure \ref{fig:mdot}a, where the transition region lies between 2.5\,G and 5\,G. 

\section{Absorption profiles for several planetary magnetic obliquities}\label{appendix:signatures}
\begin{figure*}
    \centering
    \includegraphics{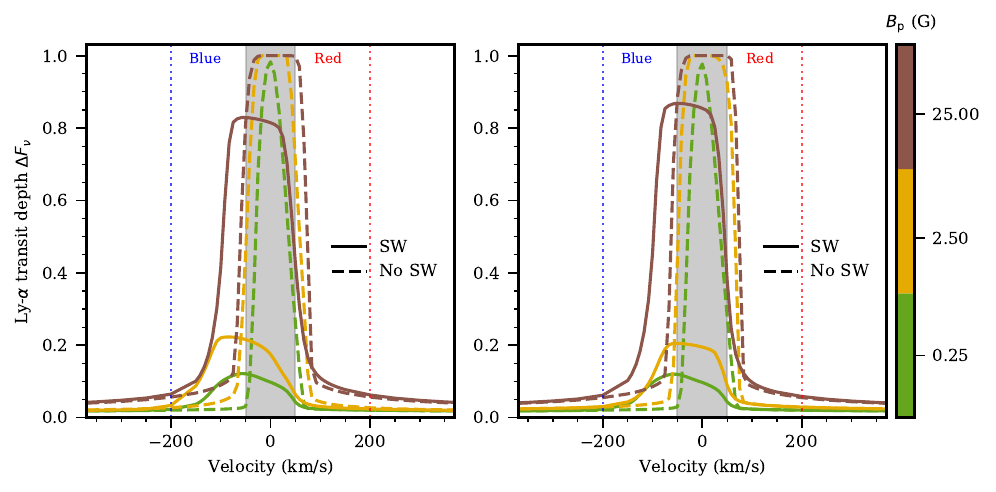}
    \caption{Similar to Figure \ref{fig:transit_depth}a, for magnetic obliquities of 45$^\circ$ (left) and 90$^\circ$ (right).}
    \label{fig:Lyman_alpha_appendix}
\end{figure*}

In Figure \ref{fig:Lyman_alpha_appendix}, we present the plots equivalent to the one considered in Figure \ref{fig:transit_depth}, but for planets with magnetic obliquities of $45^\circ$ and $90^\circ$.
Compared to the untilted cases shown in Figure \ref{fig:transit_depth}, these models display more absorption in the line centre and the blue wing, particularly for the 25\,G models with stellar wind. As discussed in section \ref{sec:spectra}, this phenomenon is caused by the position of the magnetic poles of the planet relative to the orbital plane. As the dipolar axis is tilted, the poles move closer to the orbital plane and the outflow is more directed towards the line-of-sight. 

We also note that the absorption profile for $B_{ p}=2.5$\,G appears more blueshifted in the 45$^\circ$ magnetic obliquity case compared to the other magnetic obliquities. This is because, in the $45^\circ$ case, the planet still remains in the magnetically unconfined regime for a 2.5\,G magnetic field strength, causing the tail of escaping material to be more centered in the orbital plane. 
 


\bsp	
\label{lastpage}
\end{document}